\documentclass[amsmath,amssymb,aps,prl,twocolumn]{revtex4-1}
\usepackage[dvips]{graphicx}
\usepackage{subfigure}
\usepackage{dcolumn}
\usepackage{bm}
\usepackage{hyperref}
\usepackage{array}
\usepackage{upgreek}
\usepackage[usenames]{color}
\usepackage[utf8]{inputenc}
\usepackage[T1]{fontenc}
\usepackage{mathptmx}
\usepackage{cancel}

\newcolumntype{P}[1]{>{\centering\arraybackslash}p{#1}}
\begin{document}
\preprint{AIP/123-QED}
\title{Emergent inductance by dynamical Aharonov-Casher phases}
\author{Yuta Yamane,$^{1,2}$ Shunsuke Fukami,$^{2,3,4,5,6}$ and Jun'ichi Ieda$^7$}

\affiliation{$^1$Frontier Research Institute for Interdisciplinary Sciences, Tohoku University, Sendai 980-8578, Japan}
\affiliation{$^2$Research Institute of Electrical Communication, Tohoku University, Sendai 980-8577, Japan}
\affiliation{$^3$Center for Spintronics Research Network, Tohoku University, Sendai 980-8577, Japan}
\affiliation{$^4$Center for Science and Innovation in Spintronics, Tohoku University, Sendai 980-8577, Japan}
\affiliation{$^5$Center for Innovative Integrated Electronic Systems, Tohoku University, Sendai 980-0845, Japan}
\affiliation{$^6$WPI-Advanced Institute for Materials Research, Tohoku University, Sendai 980-8577, Japan}
\affiliation{$^7$Advanced Science Research Center, Japan Atomic Energy Agency, Tokai 319-1195, Japan}
\date{\today}

\begin{abstract}

We propose a mechanism of inductance operation originating from a dynamical Aharonov-Casher (AC) phase of an electron in ferromagnets.
By taking into account spin-orbit coupling effects, we extend the theory of emergent inductance, which has recently been discovered in spiral magnets, to arbitrary magnetic textures.
The inductance of dynamical AC phase origin universally arises in the coexistence of magnetism and a spin-orbit coupling, even with spatially-uniform magnetization, allowing its stable operation in wide ranges of temperature and frequency.
Revisiting the widely studied systems, such as ferromagnets with spatial inversion asymmetry, with the new perspective offered by our work will lead to opening a new paradigm in the study of AC phase physics and the spintronics-based power management in ultra-wideband frequency range.
\end{abstract}

\maketitle

%
%
Inductors are one of the most vital electronic components in today's technology for electric voltage transformation, electric noise filtering, and so on.
Its working principle, however, has been essentially the same since its earliest inventions in 19th century;
a conducting coil stores the energy in a magnetic field when an electric current flows through it, leading to the induction of an electromotive force that opposes the change in the current.
This classical effect can also be formulated as resulting from the time derivative of Aharanov-Bohm phase, i.e., a Berry phase acquired by an electron wave function moving in an electromagnetic potential.

\begin{table*}[t]
	\centering
	\begin{tabular}{ p{4cm} P{4cm} P{4cm} P{4cm} }
		\hline
		& \multicolumn{1}{c}{Classical inductance} & \multicolumn{2}{c}{Emergent inductance} \\
		\hline \hline
		\multicolumn{1}{l}{} & &  \multicolumn{1}{c}{Spiral-based} &  \multicolumn{1}{c}{Spin-orbit} \\
		\hline
		\multicolumn{1}{l|}{System size dependence} & $ L \propto A $
	                                                                                 & $ L \propto A^{-1} $
	                                                                                 & $ L \propto A^{-1} $ \\
		\multicolumn{1}{l|}{Physical origin} & Dynamical \par Aharonov-Bohm phase
		                                                       & Dynamical \par magnetic texture
		                                                       & Dynamical \par Aharonov-Casher phase \\
		\multicolumn{1}{l|}{Characteristic frequency} & $ (LC)^{-1/2} $
		                                                                       & Pinning frequencies \par ($\sim$ MHz)
		                                                                       & Ferromagnetic resonance
		                                                                           \par frequency ($\sim$ 1-10 GHz) \\
		\multicolumn{1}{l|}{(Candidate) materials} & Coil with \par a magnetic core
		                                                                   & Gd$_3$Ru$_4$Al$_{12}$\cite{Yokouchi} \par
		                                                                       YMn$_6$Sn$_6$\cite{Kitaori}
		                                                                   & (Ga,Mn)As/GaAs\cite{Chernyshov, Fang, Kurebayashi2014} \par
		                                                                       Pt/Co/AlO$_x$\cite{Miron, Miron2011} \par
		                                                                       Ta/CoFeB/MgO\cite{Liu2012, Fukami}\\
		\hline
	\end{tabular}
	\caption{Comparison of classical and two emergent inductances in several aspects.
	              $ L $, $ A $, $ C $ represent the inductance, cross sectional area,
	              and capacitance of the corresponding inductors, respectively.}
\end{table*}

Recently, an inductance of quantum mechanical origin was predicted\cite{Nagaosa2019} and experimentally demonstrated\cite{Yokouchi,Kitaori} to arise in spiral magnets, and coined as emergent inductance.
When an electric current flows in a spiral magnet, it stores the energy in the spiral structure formed by the local magnetization via its exchange coupling with conduction electrons.
The spiral-based emergent inductance can be formulated based on a dynamical Berry phase originating from the exchange coupling, where the role of the electromagnetic potential in the classical inductance is played by a spin-dependent Berry connection generated by the spatial variation of the magnetization\cite{Berger,Volovik,Stern,Barnes}.
More recent theoretical works have shown that two excitation modes of a spiral magnetic texture, namely its translational displacement and rotation of the spiral plane, contribute to emergent inductance with opposite signs\cite{Ieda-prb,Kurebayashi}.
This may explain the negative inductance observed in the experiment\cite{Yokouchi}, whereas the original theory, which takes only into account the rotational excitation, predicted positive\cite{Nagaosa2019}.
Experimentally, a room temperature observation of emergent inductance has been achieved in YMn$_6$Sn$_6$\cite{Kitaori}, while in Ref.~\cite{Yokouchi} the temperature had to be as low as below $\sim20$ K with Gd$_3$Ru$_4$Al$_{12}$ used.

The discovery of the emergent inductance in spiral magnets has reopened the textbook of electronics, and we are at the beginning of a new chapter exploring quantum mechanical principles for inductors.
The purpose of this work is to extend the concept of emergent inductance to arbitrary magnetic textures beyond the spiral, by taking into account spin-orbit (SO) coupling effects.
We propose an emergent inductance arising due to a dynamical Aharonov-Casher (AC) phase, i.e., a Berry phase that requires the presence of SO couplings\cite{Aharonov}.
Historically, the study of AC phase has mostly been focused on its quantum interference effects on the particle transport in ring topology of mesoscopic size\cite{Cimmino,Mathur,Aronov,Konig,Bergsten}.
It has also been known since the early days that the time derivative of an AC phase can translate to a force on an electron\cite{Balatsky,Ryu}, and there is recently a resurgence of interest in a version of this dynamical effect manifested in ferromagnetic nanostructures\cite{Kim,Tatara,Yamane2013,Ciccarelli}, where the electron spin is polarized by the magnetization.
Explicit physical implications of a dynamical AC phase driven by electric current in magnetic materials, however, have never been addressed.
In Ref.~\cite{Ieda-prb}, the authors revealed complementary effects of a Rashba-type SO coupling in a spiral-based emergent inductance.
We now further generalize the concept of emergent inductance of AC phase origin, which we call ``SO emergent inductance,'' for arbitrary magnetic textures and a general form of SO couplings.
We put a particular focus on exploring SO emergent inductance with spatially uniform magnetization, where the other inductance mechanisms are ruled out.
We find that the SO emergent inductance appears in both the longitudinal and transverse (Hall) directions with respect to a current, and operates in wider ranges of temperature and frequency than the spiral-based one does.
As the SO emergent inductance turns out a rather ubiquitous phenomenon in the simultaneous presence of magnetism and a SO coupling, it sheds renewed lights on the interpretations of the existing reports on current-driven effects in those systems, such as electrical control of the magnetization exploiting SO couplings arising due to spatial inversion asymmetry.
Table I summarizes key aspects of the SO emergent inductance in comparison to other two mechanisms, which are discussed in detail in the following.

%
%
{\it General formalism ---}
We assume a conduction electron to be described by Hamiltonian $ H = \vec p^2 / 2 m_{\rm e} + J \vec \sigma \cdot \vec m + \sum_{ i j } p_i g_{ i j } \sigma_j $;
where $ m_{\rm e} $ and $ \vec p $ are the electron's effective mass and canonical momentum operator, respectively;
the second term is the exchange coupling between the conduction electron spin, represented by the Pauli matrices vector $ \vec \sigma $, and the local magnetization field, whose direction is given by the classical unit vector $ \vec m $, with the coupling constant $ J ( > 0 ) $;
the last term is a SO coupling, where the constants $ g_{ i j } $ $ ( i, j = x, y, z ) $ describes the linear coupling between the electron spin and momentum, the explicit structure of which is determined by the system symmetry.
Throughout this work, we assume $ J \gg \hbar k_{\rm F} | g_{ i j } | $ $ \forall ( i, j ) $, with $ k_{\rm F} $ denoting the Fermi wave number, so that the electron spin dynamics is predominantly dictated by the exchange coupling term.
The U(1) wave function $ | \psi^\pm \rangle $ of a conduction electron, with $ + ( - ) $ corresponding to the majority (minority) spin with respect to the magnetization direction, can then be written by
\begin{equation}
	\langle \vec r | \psi^\pm \rangle
		= \exp\left( i  \int_{\vec R}^{\vec r} d\vec r \cdot \vec a^\pm \right)
		   \langle \vec r | \psi_0^\pm \rangle ,
\end{equation}
where $ \vec R $ is some reference point in real space, $ | \psi_0^\pm \rangle $ is the wave function in the absence of SO coupling, and the Berry connection $ \vec a^\pm  = \langle \psi^\pm | i \nabla_{\vec r} | \psi^\pm \rangle $ in real space is given by
\begin{equation}
	a^\pm_i = \mp m_{\rm e} \sum_j g_{ i j } m_j ,
\label{a}
\end{equation}
with $ \nabla_{\vec r} $ denoting the derivative with respect to the vector $ \vec r $.

The time derivative of $ \vec a^\pm $ defines the spin-dependent ``electric'' field $ \vec e^\pm = - \partial_t \vec a^\pm / e $ acting on the conduction electrons, with $ e ( > 0 ) $ the elementary electric charge.
An electromotive force $ V $ arising due to $ \vec e^\pm $ in a closed circuit can be written by $ V = ( p / e ) d_t \gamma^+ $, where $ \gamma^\pm = \oint d\vec r \cdot \vec a^\pm $ is the U(1) AC phase and $ p $ is the spin polarization of the conduction electrons, which is involved because $ \vec e^\pm $ changes its sign for the majority-spin and minority-spin electrons.
The corresponding inductance of the circuit, i.e., SO emergent inductance, is then defined by $ L = - ( d_t I )^{-1} V $, where $ I $ is the electric current.
It is thus the current-driven dynamics of $ \vec a^\pm $ that is responsible for the SO emergent inductance.
Hereafter, we assume for simplicity that $ g_{ i j } $ are constant in time and thus a SO emergent inductance is generated by the directional dynamics of $ \vec m $.

Whereas  the general expression of the SO emergent inductance given above is applicable to an arbitrary dynamical magnetic texture, we focus in the following on spatially-uniform magnetization to illustrate fundamental features of the phenomenon.
Let us consider an ac electric current, with the current density $ \vec j_{\rm c} = \vec j_\omega e^{ i \omega t } $ and angular frequency $ \omega $, applied along one side of a rectangular ferromagnet to drive the magnetization dynamics $ \vec m = \vec m_0 + \vec m_\omega e^{ i \omega t } $.
Here $ \vec m_0 $ represents the time-independent part and $ \vec m_\omega $ is a complex amplitude.
In this case, the electromotive force induced along each side of the sample is simply obtained by the corresponding one-dimensional integral of the spin-dependent electric field.
The complex inductance $ L_{ i j } $ arising along the $ x_i $ direction due to the current flowing in the $ x_j $ direction is then obtained as
\begin{equation}
	L_{ i j } = \frac{ p m_{\rm e} }{ e } \frac{ l_i }{ A } \sum_k g_{ i k }  \chi_\omega^{ k j } ,
	\label{l}
\end{equation}
where $ l_i $ is the dimension of the sample in the $ x_i $ direction, $ A = | I | / | \vec j_{\rm c} | $ is the cross sectional area normal to the electric current, and $ \chi_\omega^{ i j } = m_\omega^i /  j_\omega^j $.
Equation~(\ref{l}) determines the structure of the inductance with given symmetries of $ g_{ i j } $ and $ \chi_\omega^{ i j } $.
The transverse, or Hall, inductances $ L_{ i j } $ $ ( i \neq j ) $ can appear as well as the longitudinal ones $ L_{ i i } $.
The factor $ l_i / A $ reflects the sample-dimension dependence of $ L_{ i j } $.

%
%
{\it Uniaxial ferromagnet ---}
There are rather simple cases where analytical expressions of $ \chi_\omega^{ i j } $ in terms of $ g_{ i j } $ are available.
As an example, here we examine a ferromagnet possessing a uniaxial magnetic anisotropy with the $ z $ axis set to its easy axis.
We assume that the magnetization $ \vec m $ obeys the Landau-Lifshitz-Gilbert (LLG) equation $ \partial_t \vec m = - \gamma \vec m \times ( \vec h_K + \vec h_{\rm g} ) + \alpha \vec m \times \partial_t \vec m $, where $ \gamma $ is the gyromagnetic ratio, $ \vec h_K = ( 2 K / \mu_0 M_{\rm S} ) \vec e_z $ is the effective field due to the magnetic anisotropy with $ \vec e_i $ and $ K $ being the unit vector in the $ x_i $ direction and the anisotropy constant, respectively, $ M_{\rm S} $ is the saturation magnetization, and $ \alpha $ is the Gilbert damping constant.
The vector $ \vec h_{\rm g} $ describes the effective magnetic field originating from the SO coupling, whose expression can be derived, by assuming the adiabatic approximation for the electron spin dynamics\cite{Manchon}, as
\begin{equation}
	h_{\rm g}^i = - \frac{ p m_{\rm e} }{ e \mu_0 M_{\rm S} } \sum_j g_{ j i } j_{\rm c}^j .
\label{hg}
\end{equation}
The underlying physics of $ \vec h_{\rm g} $ is understood as the SO coupling in the Hamiltonian $ H $ causes a slight deviation of the electron spin from $ \vec m $, which in turn exerts on $ \vec m $ the momentum-dependent effective field~(\ref{hg}) via the exchange coupling.

We assume the electric current to be applied in the $ x $ direction and, for the sake of simplicity, that $ g_{ i j } = 0 $ if either $ i $ or $ j $ is $ z $.
The latter assumption holds for most SO couplings that are of practical importance\cite{Winkler,Manchon2015}.
The effective field $ \vec h_g $ thus lies in the $ x y $ plane.
The LLG equation can now be linearized by employing a small cone-angle ansatz $ \vec m \simeq m_0 \vec e_z + e^{ i \omega t } ( m_\omega^x \vec e_x + m_\omega^y \vec e_y ) $, with $ m_0 = \pm 1 $ and $ | m_\omega^{ x, y } | \ll 1 $.
The dynamical susceptibilities $ \chi_\omega^{ i x } ( i = x, y ) $ are obtained as
\begin{eqnarray}
	\left( \begin{array}{c} \chi_\omega^{ x x }  \\  \chi_\omega^{ y x } \end{array} \right)
		&=& \frac{ 1 }{ \omega_K^2 - ( 1 + \alpha^2 ) \omega^2
		                   + 2 i \alpha \omega_K \omega } \nonumber \\ && \times
		   \left( \begin{array}{cc} \omega_K + i \alpha \omega & i \omega m_0 \\
	                                               - i \omega m_0 & \omega_K + i \alpha \omega
	           \end{array} \right)
	           \left( \begin{array}{c} u_x \\ u_y \end{array} \right) ,
\label{chi}
\end{eqnarray}
where $ \omega_K = \gamma h_K $ and $ \vec u = \gamma \vec h_{\rm g} / j_\omega^x $.
Combining Eq.~(\ref{l}) with (\ref{hg}) and (\ref{chi}), the longitudinal and Hall inductances, $ L_{ xx} $ and $ L_{ yx} $, respectively, are obtained.

With the ferromagnetic resonance frequency $ \omega_{\rm R} $ defined by $ \omega_{\rm R} = \omega_K ( 1 + \alpha^2 )^{ - 1/2 } $, the dynamical susceptibility can be approximated by $ (  \chi_\omega^{ x x } ,  \chi_\omega^{ y x } ) \simeq - \frac{ p m_{\rm e} }{ 2 e K } ( g_{ x x} , g_{ x y } ) $  for $ \omega \ll \omega_{\rm R} $.
In this low frequency regime, the inductances can be reduced to
\begin{eqnarray}
	L_{ x x } &=& \left( \frac{ p m_{\rm e} }{ e } \right)^2 \frac{ l_x }{ A }
	                     \frac{ g_{ x x}^2 + g_{ x y }^2 }{ 2 K } , \label{l0} \\
	L_{ y x } &=& \left( \frac{ p m_{\rm e} }{ e } \right)^2 \frac{ l_y }{ A }
	                       \frac{ g_{ y x } g_{ x x } + g_{ y y } g_{ x y } }{ 2 K } . \label{ly}
\end{eqnarray}
Note that all the terms in $ L_{ x x }$ and $ L_{ y x } $ are second order of $ g_{ i j } $.
This is because the SO emergent inductance is a result of a sequential action of the current-induced torque on the magnetization and the spin-dependent electric field driven by the magnetization dynamics, each of the two effects being linear in $ g_{ i j } $, see Eqs.~(\ref{a}) and (\ref{hg}).

So far, we have discussed the SO emergent inductance within the adiabatic limit for the electron spin dynamics.
Nonadiabatic corrections can be shown to give rise to the additional spin-dependent electric field $ e^{ \beta \pm}_i = \mp \frac{ \beta m_{\rm e} }{ e } \sum_j g_{ i j } ( \vec m \times \partial_t \vec m )_j  $ \cite{Tatara,Yamane2013} and effective magnetic field $ \vec h^\beta_{\rm g} = \beta \vec m \times \vec h_{\rm g} $ \cite{Kim2012,Pesin}, where $ \beta ( \ll 1 ) $ is a dimensionless constant that phenomenologically characterizes the nonadiabaticity in the electron spin dynamics.
There arise additional inductances due to $ \vec e^{ \beta \pm } $ and $ \vec h_{\rm g}^\beta $.
With the same ansatz for the magnetization dynamics as used in deriving Eqs.~(\ref{l0}) and (\ref{ly}), the nonadiabatic inductances for $ \omega \ll \omega_{\rm R} $ are obtained by
\begin{eqnarray}
	L_{ x x }^\beta &=& - \beta^2 L_{ x x } , \\
	L_{ y x }^\beta &=& - \beta^2 L_{ y x } +  \left( \frac{ p m_{\rm e} }{ e } \right)^2 \frac{ l_y }{ A }
	                                                                 \frac{ \beta m_0 \left( g_{ x y } g_{ y x } - g_{ x x } g_{ y y } \right) }{ K } .
\label{lb}
\end{eqnarray}
Note that, since the second term in Eq.~(\ref{lb}) contains $ m_0 $, it changes the sign as the equilibrium magnetization direction is reversed between $ + \vec e_z $ and $ - \vec e_z $.

\begin{figure}[t]
  \centering
  \includegraphics[width=7cm, bb=0 0 1394 2157]{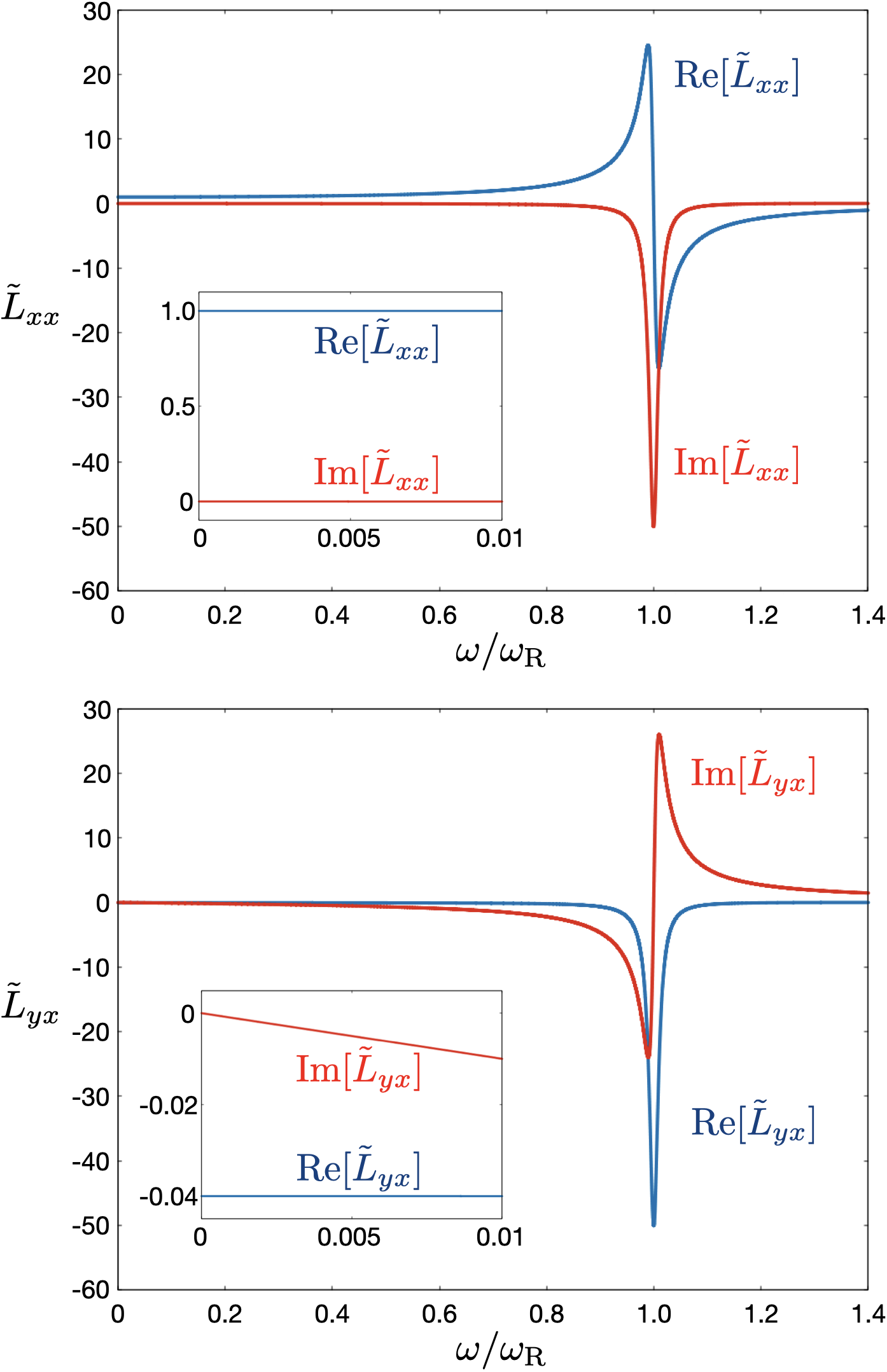}
  \caption{ Frequency characteristics of the longitudinal ($ \tilde { L }_{ x x } $) and Hall ($ \tilde{ L }_{ y x } $) spin-orbit emergent inductances, normalized by the factors associated with their low frequency values (see the main text).
   Here we assume $ m_0 = + 1 $, $ g_{ y x } = - g_{ x y } = {\rm const}. $, $ g_{ i j } = 0 $ for the other $ ( i , j ) $, and $ \beta = 0.02 $;
  for the reversed magnetization $ m_0 = - 1 $, $ \tilde{ L }_{ x x } $ is identical with (a), while $ \tilde{ L }_{ y x } $ is obtained by reversing the sign of the curves in (b).
The insets are magnified views of the low frequency regime.
                               }
  \label{fig01}
\end{figure}

Equations~(\ref{l0})-(\ref{lb}) indicate that, for $ \omega \ll \omega_{\rm R} $, both the longitudinal and Hall inductances are real and independent of $ \omega $.
Plotted in Fig.~1 are the real and imaginary parts of the inductances based on Eq.~(\ref{chi}) as functions of $ \omega $, including the vicinity of $ \omega_{\rm R} $.
Here we assume a Rashba-type SO coupling with $ g_{ y x } = - g_{ x y } \equiv g $ (constant) and $ g_{ i j } = 0 $ for the other $ ( i , j ) $.
The normalized inductances are defined by $ \tilde{ L }_{ x x } \equiv ( L_{ x x } + L_{ x x }^\beta ) / L_x^0 $ and $ \tilde{ L }_{ y x } \equiv ( L_{ y x } + L_{ y x }^\beta ) / L_y^0 $, with the normalization factors $ L^0_i = ( p m_{\rm e} g / e )^2 ( l_i / A ) / 2 K $.
The enhancement in the magnitude of inductances is observed at around $ \omega_{\rm R} $, in parallel with that of $ \chi_\omega^{ x x } $ and $ \chi_\omega^{ y x } $, and the sign of $ {\rm Re} [ \tilde{ L }_{ x x } ] $ and $ {\rm Im} [ \tilde{ L }_{ y x } ] $ is reversed across $ \omega_{\rm R} $.
Shown in the insets are magnified views of the low frequency regime, where the real parts hardly exhibit the $ \omega $ dependence and the imaginary parts are relatively small, being consistent with Eqs.~(\ref{l0})-(\ref{lb}).
With the present structure of $ g_{ i j } $, it can be shown that $ \tilde{ L }_{ y x } $ is overall proportional to $ m_0 $ while $ \tilde{ L }_{ x x } $ is independent of the same factor.
This is also the case for a Dresselhaus-type coupling, another common type of SO coupling, with $ g_{  x x } = - g_{ y y } $ and $ g_{ i j } = 0 $ for the other $ ( i, j ) $.

%
%
{\it Discussion ---}
From technological application points of view, the emergent inductance offers an advantage over the classical counterpart in nanoscale device fabrications.
The former is inversely proportional to the system's cross sectional area (normal to the electric current direction)\cite{Nagaosa2019}, which is in a strike contrast with the latter being directly proportional to the cross sectional area of a coil.
Ref.~\cite{Yokouchi} reported an emergent inductance of $ \sim 400 $ nH, which is comparable in its magnitude to that of a commercial one, but in a volume about a million times smaller.
The spiral-based emergent inductance confronts, however, its own issue, which is the limitation in the operating frequency.
A spiral magnetic structure cannot respond collectively and robustly to electric currents with higher frequency than the characteristic value determined by the pinning potentials for spiral dynamics.
The reported experiments successfully observed the emergent inductance up to the frequencies of sub-megahertz\cite{Kitaori} and megahertz\cite{Yokouchi}.
 The SO emergent inductance resolves simultaneously the two issues regarding miniaturization and the operating frequency.
As seen in Eq.~(\ref{l}), the smaller the cross sectional area $ A $, the larger the SO emergent inductances, as with the case for the spiral-based one.
The SO emergent inductance with spatially uniform magnetization, as shown in Fig.~1, also provides the nearly $ \omega$-independent real parts except at the vicinity of the resonance frequency $ \omega_{\rm R} $, which is typically $\sim$~1-10~GHz.

Potential candidate systems for experimental demonstration of the SO emergent inductance include ferromagnets with spatial inversion asymmetry, where Rashba and Dresselhaus SO couplings can arise\cite{Winkler}.
Those SO couplings have been extensively studied as sources of the effective magnetic fields acting on semiconducting \cite{Chernyshov, Fang, Kurebayashi2014} as well as metallic\cite{Miron, Miron2011} ferromagnets.
Because the SO emergent inductance shows up below the ferromagnetic transition temperature, the metallic systems are particularly promising platforms for its room temperature observation.
The magnitude of $ g_{ i j } $ largely varies depending on the system, and their accurate estimation has been a subject of debate.
Adopting here $ g = 10^{-10} $ eV$\cdot$m/$\hbar$ \cite{Miron}, and employing some typical values for the other relevant material parameters as $ K \sim 10^5$ J/m$^3$, $ p \sim 0.5 $, and the bare electron mass for $ m_{\rm e} $, $ L_{ x x } $ in Eq.~(\ref{l0}) is estimated as $ \sim 10^{-18} \times ( l_x / A ) $~H.
Assuming the sample dimensions of $ ( l_x, l_y, l_z ) = ( 0.1{\rm mm}, 100{\rm nm}, 10{\rm nm} ) $, we thus arrive at an estimation $ L_{ x x } \sim 100 $ nH, which is comparable to the recently reported spiral-based emergent inductance\cite{Yokouchi}.
(Similar estimations for Eqs.~(\ref{ly})-(\ref{lb}) with the same set of parameters are obvious.)
In the context of the conventional spintronics applications, larger $ K $ has been mostly pursued for the sake of better thermal stability of information stored in the magnetization profile.
For a larger SO emergent inductance, in contrast, smaller $ K $ is rather preferred, see Eqs.~(\ref{l0})-(\ref{lb}), while it also leads to a lower resonance frequency.
It is thus desired to systematically conduct experimental and material research in the future to achieve optimal conditions for the SO emergent inductance, i.e., simultaneous realization of larger $ g_{ i j } $, smaller $ A $, and the right magnitude of $ K $ depending on the purpose.

We started from the particular type of Hamiltonian $ H $ for the clarity of discussion, but the presence of the inductance of SO coupling origin does not depend on the specific model.
In ferromagnet/nonmagnet heterostructures, it is known that the spin Hall effect arising due to the atomic SO coupling also contributes effective magnetic fields\cite{Kim2013, Garello} independently of the SO couplings discussed above, showing promise for non-volatile memory applications\cite{Liu2012, Fukami, Shao}.
The induced magnetization dynamics then produces an electromotive force via the inverse spin Hall effect\cite{Saitoh}.
From symmetry argument\cite{Manchon2019}, this process can be seen to generate an inductance in the same way as the Rashba SO coupling does, potentially making some impacts on the circuit operation.

In summary, we have extended the theory of emergent inductance, deriving a novel inductance of dynamical AC phase origin.
The proposed inductance is ubiquitously expected in wide ranges of temperature and frequency when magnetism and a SO coupling coexist.
While electrical control of the magnetization in those systems utilizing the effective fields has become a central research subject in spintronics, the accompanying effects originating from the dynamical AC phase have been largely overlooked.
Our findings suggest that those spintronics devices should be designed with the proposed dynamical AC phase effect in their circuit operation taken into account.
A key to an exploration of this new physical implication of an AC phase and an innovative advance in inductor technologies is to revisit the extensively studied systems, such as ferromagnets with spatial inversion asymmetry, with the new perspective of emergent inductance.


{\it Acknowledgments ---}
The authors wish to thank S. Kanai, H. Ohno, K. Yamamoto and Y. Araki for fruitful discussions and their valuable comments on the work.
This work was supported by Iketani Science and Technology Foundation, Murata Science Foundation, and JSPS KAKENHI, Grant No. 16K18084, JP16K05424, 19J12926, JP19H05622, and 20J14418.

{}

\end{document}